\documentclass
[superscriptaddress,secnumarabic,amssymb,amsmath,nobibnotes,aps,prd,showkeys,showpacs,nofootinbib,onecolumn,notitlepage,12pt]{revtex4}%
\usepackage{setspace}
\usepackage{color}
\usepackage{amsmath}
\usepackage{amsfonts}
\usepackage{verbatim}
\usepackage{amssymb}
\usepackage{graphicx,bm}
\usepackage{amsmath}
\usepackage{amssymb}
\usepackage{graphicx}
\usepackage[caption=false]{subfig}%
\providecommand{\U}[1]{\protect\rule{.1in}{.1in}}

\newcommand{\be}{\begin{equation}}
\newcommand{\ee}{\end{equation}}

\newcommand{\mincir}{\raise
-3.truept\hbox{\rlap{\hbox{$\sim$}}\raise4.truept\hbox{$<$}\ }}
\newcommand{\magcir}{\raise
-3.truept\hbox{\rlap{\hbox{$\sim$}}\raise4.truept\hbox{$>$}\ }}

\ifx\pdfoutput\relax\let\pdfoutput=\undefined\fi
\newcount\msipdfoutput
\ifx\pdfoutput\undefined\else
\ifcase\pdfoutput\else
\ifx\paperwidth\undefined\else
\ifdim\paperheight=0pt\relax\else\pdfpageheight\paperheight\fi
\ifdim\paperwidth=0pt\relax\else\pdfpagewidth\paperwidth\fi
\fi\fi\fi
\begin{document}
\title{Noether Symmetry Analysis for Generalized Brans-Dicke Cosmology}
\author{Andronikos Paliathanasis}
\email{anpaliat@phys.uoa.gr}
\affiliation{Institute of Systems Science, Durban University of Technology, Durban 4000,
South Africa}
\affiliation{Centre for Space Research, North-West University, Potchefstroom 2520, South Africa}
\affiliation{Departamento de Matem\'{a}ticas, Universidad Cat\'{o}lica del Norte, Avda.
Angamos 0610, Casilla 1280 Antofagasta, Chile}
\affiliation{School of Sciences, Woxsen University, Hyderabad 502345, Telangana, India}

\begin{abstract}
We present the complete solution to the classification problem regarding the
variational symmetries of the generalized Brans-Dicke cosmological model in
the presence of a second scalar field minimally coupled to gravity and the
generalized Brans-Dicke scalar field theory. Through the symmetry analysis, we
were able to specify the functional form of the field equations such that they
become integrable. Additionally, new families of integrable cosmological
models are presented.

\end{abstract}
\maketitle
\date{\today}

\section{Introduction}

\label{sec1}

Among the many models suggested by cosmologists to address the puzzle of
cosmological observations \cite{Teg,Kowal,Komatsu,suzuki,cosm1}, scalar fields
play a crucial role \cite{ten2,ten3,ten4,ten5,ten6,ten7,ten8}. A scalar field
model that has drawn attention is the Brans-Dicke theory \cite{Brans} and its
modifications \cite{faraonibook}. These theories, known as scalar-tensor
gravitational models, introduce a coupling function between the scalar field
and the gravitational Lagrangian, which means that the existence of the scalar
field is essential for the appearance of the gravitational field.
Scalar-tensor theories satisfy Mach's principle \cite{mc1,mc2}. Furthermore,
the scalar-tensor theories belong to the family of Horndeski models
\cite{sa1}, which means that the gravitational field equations are of second
order and do not suffer from Ostrogradsky's instabilities.

On large scales, the universe is assumed to be described by the isotropic and
homogeneous Friedmann--Lema\^{\i}tre--Robertson--Walker (FLRW) spacetime.
Despite the fact that the gravitational field equations in scalar-tensor
theories are of second order, and the number of dynamical degrees of freedom
for the gravitational field equations in cosmology is relatively small, there
are very few exact and analytic cosmological solutions in the literature; see,
for instance, \cite{sol1,sol2,sol3,sol4} and references therein. There exists
a minisuperspace description for the gravitational field equations in
cosmology. Therefore, there exists a point-like Lagrangian which describes the
field equations.

With this property in mind, in \cite{annoe1}, Noether's theorems were applied
in order to constrain the free parameters of the scalar-tensor theory so that
variational symmetries could exist. By applying Noether's second theorem,
conservation laws were determined, and the field equations were solved
explicitly \cite{ns1}. By imposing the constraint equation, a similar analysis
was performed in \cite{ndim1}. The Noether symmetry analysis was extended to
the case of Brans-Dicke theory in the presence of a matter source with a
constant equation of state parameter \cite{annoe2} and in the presence of a
second scalar field \cite{annoe3}. The application of Noether symmetry
analysis in gravitational physics leads to the Ovsiannikov group
classification problem \cite{Ovsi}, where all the free parameters of the model
are constrained so that all the possible admissible symmetries can be
calculated. For more details, we refer the reader to the discussion in
\cite{ibra,olver}. Moreover, as has been discussed in \cite{annoe2}, the
Noetherian conservation laws can be used to define quantum observables which
can be used to solve the Wheeler-DeWitt equation of quantum cosmology.

Recently, in \cite{com1}, the Noether symmetry analysis was applied to the
study of the classical and quantum cosmological equations for a generalized
Brans-Dicke model with an additional scalar field to describe the matter
source. Nevertheless, the analysis presented in \cite{com1} is incomplete, and
the authors have not solved Ovsiannikov's group classification problem related
to the symmetry analysis.

In this study, we present the complete classification of the Noether symmetry
analysis for the same generalized Brans-Dicke cosmological model with a scalar
field. We show that new analytic solutions can be derived, relating to
integrable Hamiltonian systems for specific forms of the free functions of the
theory. The structure of the paper is as follows.

In Section \ref{sec2}, we present the gravitational model under our
consideration, which is a scalar-tensor theory where the matter source
component is described by a scalar field minimally coupled to gravity. The
case of a FLRW geometry is discussed, and the basic properties of the
variational symmetries are presented. In Section \ref{sec3}, we present the
solution to the Noether symmetry classification problem for the cosmological
model under our consideration. We find that the existence of symmetries
depends on four different coupling functions for the nonminimally coupled
scalar field and on four different potential functions for the second scalar
field. This analysis shows that the results published in \cite{com1} are only
a fraction of the complete results. Furthermore, in Section \ref{sec4}, we
employ the Noether symmetries to determine the integrability properties and to
reduce the dynamical systems to other known dynamical systems that are
integrable. Finally, in Section \ref{sec5}, we draw our conclusions.

\section{Generalized Brans-Dicke cosmology}

\label{sec2}

We consider the scalar-tensor theory expressed by the Action Integral
\cite{faraonibook}%
\begin{equation}
S=\int dx^{4}\sqrt{-g}\left[  \frac{1}{2}\phi R-\frac{1}{2}\frac{\omega
_{BD}\left(  \phi\right)  }{\phi}g^{\mu\nu}\phi_{;\mu}\phi_{;\nu}-L_{m}\left(
g_{\mu\nu},\psi,\psi_{;\mu}\right)  \right]  ,\label{ac.01}%
\end{equation}
in which $\phi\left(  x^{\kappa}\right)  $ is the scalar field nonnominally
coupled to gravity, $\omega_{BD}\left(  \phi\right)  $, with $\omega
_{BD}\left(  \phi\right)  \neq\frac{3}{2}$, defines the coupling between the
gravity and the scalar field, for $\omega_{BD}\left(  \phi\right)  =const$,
the Brans-Dicke theory is recovered and $L_{m}\left(  g_{\mu\nu},\psi
,\psi_{;\mu}\right)  $ is the Lagrangian function of a scalar field~$\psi
\left(  x^{\kappa}\right)  $ minimally coupled to gravity and to the
Brans-Dicke scalar field, that is,%
\begin{equation}
L_{m}\left(  g_{,\mu\nu},\psi,\psi_{;\mu}\right)  =\frac{1}{2}g^{\mu\nu}%
\psi_{;\mu}\psi_{;\nu}+V\left(  \psi\right)  .\label{ac.02}%
\end{equation}
Despite the fact that function $\omega_{BD}\left(  \phi\right)  $ is the
coefficient of the kinetic term of the scalar field $\phi$, we can always
define a new scalar field $\Phi$, such that $\sqrt{\omega_{0}}d\Phi=\int
\sqrt{\frac{\omega_{BD}\left(  \phi\right)  }{\phi}}d\phi$, in order to write
the gravitational Action Integral in the equivalent form%
\begin{equation}
S=\int dx^{4}\sqrt{-g}\left[  \frac{1}{2}\phi\left(  \Phi\right)
R-\frac{\omega_{0}}{2}g^{\mu\nu}\Phi_{;\mu}\Phi_{;\nu}-L_{m}\left(  g_{,\mu
\nu},\Phi,\psi,\psi_{;\mu}\right)  \right]  ,\label{ac.03}%
\end{equation}
in which the Brans-Dicke action is recovered when $\phi\left(  \Phi\right)
=\phi_{0}\Phi^{2}$. \qquad

Variations of the Action\ Integral (\ref{ac.01}) with respect to the metric
tensor leads to the modified Einstein field equations
\begin{equation}
G_{\mu\nu}=\frac{\omega_{BD}}{\phi^{2}}\left(  \phi_{;\mu}\phi_{;\nu}-\frac
{1}{2}g_{\mu\nu}g^{\kappa\lambda}\phi_{;\kappa}\phi_{;\lambda}\right)
-\frac{1}{\phi}\left(  g_{\mu\nu}g^{\kappa\lambda}\phi_{;\kappa\lambda}%
-\phi_{;\mu}\phi_{;\nu}\right)  -g_{\mu\nu}\frac{V\left(  \phi\right)  }{\phi
}+\frac{1}{\phi}T_{\mu\nu}, \label{bd.02}%
\end{equation}
where $T_{\mu\nu}$ describes the contribution of the field $\psi\left(
x^{k}\right)  $ in the field equations, that is,%
\begin{equation}
T_{\mu\nu}=\psi_{;\mu}\psi_{;\nu}-\frac{1}{2}g_{\mu\nu}\left(  g^{\kappa
\lambda}\phi_{;\kappa}\phi_{;\lambda}+V\left(  \psi\right)  \right)  .
\label{bd.03}%
\end{equation}

Furthermore, the variation of the action integral (\ref{ac.01}) with respect
to the scalar fields $\phi\left(  x^{\kappa}\right)  ,~\psi\left(  x^{\kappa
}\right)  $ leads to the Klein-Gordon equations%

\begin{subequations}
\begin{equation}
g^{\kappa\lambda}\phi_{;\kappa\lambda}-\frac{1}{2\phi}g^{\mu\nu}\phi_{;\mu
}\phi_{;\nu}+\frac{\phi}{2\omega_{BD}}\left(  R-2V_{,\phi}\right)  =0,
\label{bd.02a}%
\end{equation}%
\end{subequations}
\begin{equation}
g^{\kappa\lambda}\psi_{;\kappa\lambda}+V_{,\psi}=0. \label{bd.02b}%
\end{equation}

\subsection{FLRW Cosmology}

For the spatially flat FLRW line element
\begin{equation}
ds^{2}=-dt^{2}+a^{2}\left(  t\right)  \left(  dx^{2}+dy^{2}+dz^{2}\right)
,\label{bd.05}%
\end{equation}
where $a\left(  t\right)  $ is the scale factor of the universe and
$H=\frac{\dot{a}}{a}$ is the Hubble function, with $\dot{a}=\frac{da}{dt}$.
The Ricci scalar is calculated.
\begin{equation}
R=6\left[  \frac{\ddot{a}}{a}+\left(  \frac{\dot{a}}{a}\right)  ^{2}\right]
.\label{bd.06}%
\end{equation}
Moreover, we assume that the scalar fields inherit the symmetries of the
background space. As a result, the field equations (\ref{bd.02}) and the
equations of motion for the scalar fields (\ref{bd.02a}), (\ref{bd.02b}) form
the following set of ordinary differential equations%

\begin{equation}
3\phi H^{2}+3H\dot{\phi}+\frac{\omega_{BD}\left(  \phi\right)  }{2\phi}%
\dot{\phi}^{2}+\frac{1}{2}\dot{\psi}^{2}+V\left(  \psi\right)  =0,
\label{cc.01}%
\end{equation}%
\begin{equation}
3\phi H^{2}+2\phi\dot{H}+2H\dot{\phi}-\frac{\omega_{BD}\left(  \phi\right)
}{2}\left(  \frac{\dot{\phi}}{\phi}\right)  ^{2}+\ddot{\phi}-\frac{1}{2}%
\dot{\psi}^{2}+V\left(  \psi\right)  =0,
\end{equation}%
\begin{equation}
\frac{\ddot{\phi}}{\phi}+\left(  \ln\left(  \frac{\omega_{BD}\left(
\phi\right)  }{\phi}\right)  \right)  _{,\phi}\frac{\dot{\phi}^{2}}{\phi
}+3H\frac{\dot{\phi}}{\phi}+\frac{1}{\omega_{BD}\left(  \phi\right)  }\left(
\left(  3\dot{H}+6H^{2}\right)  +V_{,\phi}\right)  =0,
\end{equation}%
\begin{equation}
\ddot{\psi}+3H\dot{\psi}+V_{,\psi}\left(  \psi\right)  =0.
\end{equation}

The aforementioned dynamical system follows from the variation of the
point-like Lagrangian function.
\begin{equation}
L\left(  a,\dot{a},\phi,\dot{\phi},\psi,\dot{\psi}\right)  =\frac{1}{2}\left(
-6a\phi\dot{a}^{2}-6a^{2}\dot{a}\dot{\phi}-\frac{\omega_{BD}\left(
\phi\right)  }{\phi}a^{3}\dot{\phi}^{2}-a^{3}\dot{\psi}^{2}\right)
+a^{3}V\left(  \psi\right)  \label{bd.07}%
\end{equation}
where the constraint equation (\ref{cc.01}) is the energy related to the
dynamical system with Lagrangian (\ref{bd.07}).

In the following lines we employ Noether's theorems to determine the
variational symmetries for the Action Integral defined by the Lagrangian
function (\ref{bd.07}). Specifically, the free functions $\omega_{BD}\left(
\phi\right)  $ and $V\left(  \psi\right)  $ will be constrained with the
requirement of the existence of variational symmetries.

\subsection{Variational Symmetries}

In the early twentieth century, Emmy Noether published her pioneering work on
variational symmetries and conservation laws \cite{noe0}. Noether's first
theorem provides a simple algebraic relation that can be used to calculate the
one-parameter point transformations that leave invariant the variation of the
action integral. Furthermore, Noether's second theorem relates the variational
symmetries to the admitted conservation laws for the given dynamical system.
Noether's work belongs to the wider field of Lie symmetry analysis of
differential equations. The basic elements and definitions of the Noether
symmetry analysis are presented below.

Consider the infinitesimal one parameter point transformation \cite{noe1}%
\begin{align}
\bar{t} &  =t+\varepsilon\xi\left(  t,a,\phi,\psi\right)  ,\\
\bar{a} &  =a+\varepsilon\eta_{a}\left(  t,a,\phi,\psi\right)  ,\\
\bar{\phi} &  =\phi+\varepsilon\eta_{\phi}\left(  t,a,\phi,\psi\right)  ,\\
\bar{\psi} &  =\psi+\eta_{\psi}\left(  t,a,\phi,\psi\right)
\end{align}
in which $\varepsilon$ is an infinitesimal parameter, i.e. $\varepsilon
^{2}\rightarrow0$, and $\xi,~\eta_{a},~\eta_{\phi}$ are the components of the
generator $X$ for the infinitesimal transformation defined as%
\begin{equation}
X=\frac{\partial\bar{t}}{\partial\varepsilon}\partial_{t}+\frac{\partial
\bar{a}}{\partial\varepsilon}\partial_{a}+\frac{\partial\bar{\phi}}%
{\partial\varepsilon}\partial_{\phi}+\frac{\partial\bar{\psi}}{\partial
\varepsilon}\partial_{\psi}.
\end{equation}

The variation of the Action Integral~%
\[
S=\int L\left(  a,\dot{a},\phi,\dot{\phi},\psi,\dot{\psi}\right)  dt\,
\]
$\ $is invariant if and only there exist a function $f$, such that the
following condition is true \cite{noe1,noe2}
\begin{equation}
X^{\left[  1\right]  }L+L\dot{\xi}=\dot{f},\label{Lie.5}%
\end{equation}
where $X^{\left[  1\right]  }$ is the first extension of $X$ in the jet space
defined as
\begin{equation}
X^{\left[  1\right]  }=X+\left(  \dot{\eta}_{a}-\dot{a}\dot{\xi}\right)
\partial_{\dot{a}}+\left(  \dot{\eta}_{\phi}-\dot{\phi}\dot{\xi}\right)
\partial_{\dot{\phi}}+\left(  \dot{\eta}_{\dot{\psi}}-\dot{\psi}\dot{\xi
}\right)  \partial_{\dot{\psi}}.
\end{equation}
The function $f$ is a boundary term introduced to allow for the infinitesimal
changes in the value of the Action Integral generated by the infinitesimal
change in the boundary of the domain due to the transformation.

The symmetry condition (\ref{Lie.5}) in Noether's first theorem provides a set
of constraint equations on the components of the generator $X$ and the free
parameters of the Lagrangian function. For the model under our consideration,
the scalar field potential $V\left(  \psi\right)  $ and the coupling parameter
$\omega_{BD}\left(  \phi\right)  $ are constrained such that the symmetry
condition (\ref{Lie.5}) leads to the existence of nontrivial symmetry vectors.

Noether's second theorem, relates the variational symmetries $X$ with the
conservation laws for the equations of motions. Indeed, if $X$ is a symmetry
vector the function \cite{noe1,noe2}
\begin{equation}
I\left(  t,a,\dot{a},\phi,\dot{\phi},\psi,\dot{\psi}\right)  =\xi
\mathcal{H}-\left(  \frac{\partial L}{\partial\dot{a}}\eta_{a}+\frac{\partial
L}{\partial\dot{\phi}}\eta_{\phi}+\frac{\partial L}{\partial\dot{\psi}}%
\eta_{\psi}\right)  +f
\end{equation}
is a conserved quantity, i.e. $\frac{dI}{dt}=0$; where $\mathcal{H}$ is the
Hamiltonian function defined as%
\begin{equation}
\mathcal{H}=\frac{\partial L}{\partial\dot{a}}\dot{a}+\frac{\partial
L}{\partial\dot{\phi}}\dot{\phi}+\frac{\partial L}{\partial\dot{\psi}}%
\dot{\psi}-L.
\end{equation}
Recall that for in our consideration, the Hamiltonian is zero due to the
constraint (\ref{cc.01}), that is, $\mathcal{H}=0$. The conserved quantities
related to the Noether symmetries in Newtonian Mechanics are related with
physical observables. Nevertheless till now there is a not a clear physical
meaning on the conservation laws for the field equations of gravitational
physics. Mainly these conserved quantities are used to identify integrable
systems and derive analytic solutions.

At this point it is important to mention that in this study we employ the
Noether symmetry analysis by assuming that the dynamical system is a regular
system. Nevertheless, in gravitational physics due to the constraint equation
the dynamical system can be seen as singular, and that leads to an alternative
approach to the Noether symmetry analysis as discussed in \cite{ndim1}.

We proceed with the presentation of the solution for the Noether symmetry
classification problem.

\section{Noether symmetry classification problem}

\label{sec3}

For the Lagrangian function (\ref{bd.07}) the Noether symmetry conditions
(\ref{Lie.5}) are presented in Appendix \ref{app1}. For arbitrary functional
forms of $\omega_{BD}\left(  \phi\right)  $ and $V\left(  \psi\right)  $, the
symmetry conditions implies that there is always a Noether symmetry
$X^{0}=\partial_{t}$, with the corresponding conservation law being the
Hamiltonian function $H$. This symmetry vector is trivial and indicates that
the gravitational field equations form an autonomous dynamical system.

Nevertheless, for specific functional forms of $\omega_{BD}\left(
\phi\right)  $ and $V\left(  \psi\right)  $, the cosmological model under our
consideration may admit additional nontrivial symmetries. These symmetries are
presented in the following lines.

\subsection{Arbitrary coupling function $\omega_{BD}\left(  \phi\right)  $}

For arbitrary coupling functions $\omega_{BD}\left(  \phi\right)  $, there are
three potential functions for which additional variational symmetries exist:
the zero potential $V_{A}\left(  \psi\right)  = 0$, the constant (non-zero)
potential $V_{B}\left(  \psi\right)  = V_{0}$, and the exponential potential
$V_{C}\left(  \psi\right)  = V_{0}e^{-\lambda\psi}$.

The zero potential function~$V_{A}\left(  \psi\right)  $, leads to the
appearance of the two symmetry vectors%
\[
X^{1}=\partial_{\psi}~,~X^{2}=2t\partial_{t}+\frac{2}{3}a\partial_{a},
\]
with corresponding conservation laws%
\begin{equation}
I^{1}\left(  X^{1}\right)  =\frac{3}{2}a^{3}\dot{\psi},
\end{equation}%
\begin{equation}
I^{2}\left(  X^{2}\right)  =3a^{3}\phi\left(  2H+\frac{\dot{\phi}}{\phi
}\right)  ,
\end{equation}

Furthermore, for the constant potential $V_{B}\left(  \psi\right)  $, there
exist only the additional Noether symmetry $X^{1}$, with conservation law the
function $I^{1}\left(  X^{1}\right)  $. \ 

Finally, for the exponential potential $V_{C}\left(  \psi\right)  $, the
admitted Noether symmetry is the vector field $X^{1+2}=X^{2}+\frac{4}{\lambda
}X_{1}$, with conservation law%
\begin{equation}
I^{1+2}\left(  X^{1+2}\right)  =I^{2}\left(  X^{2}\right)  +\frac{4}{\lambda
}I^{1}\left(  X^{1}\right)  .
\end{equation}

\subsection{Brans-Dicke theory}

For the case where $\omega_{BD}\left(  \phi\right)  = \omega_{BD}^{0}$, the
scalar-tensor field reduces to the Brans-Dicke theory. The Noether symmetry
classification scheme for this problem was solved previously in \cite{annoe3}
for a more general potential function that depends on the two scalar fields
and leads to the introduction of a coupling function.

It has been found that the only new potential function where a nontrivial
variational symmetry exists is the power-law potential $V_{D}\left(
\psi\right)  = V_{0} \left(  \psi- \psi_{0} \right)  ^{2 + \alpha}$, with the
Noether symmetry associated with the vector field%

\begin{equation}
X^{3}=\alpha X^{2}+\frac{8}{3}\left(  -a\partial_{a}+3\phi\partial_{\phi
}+\frac{3}{2}\left(  \psi-\psi_{0}\right)  \partial_{\psi}\right)  ,
\end{equation}
with Noetherian conservation law%
\begin{equation}
I^{3}\left(  X^{3}\right)  =I^{2}\left(  X^{2}\right)  +\frac{8}{3}\left[
3a^{2}\dot{\phi}a-12a^{2}\dot{a}\phi-\frac{3\omega_{BD}^{0}}{\phi}a^{3}%
\phi\dot{\phi}-\frac{3}{2}a^{3}\left(  \psi-\psi_{0}\right)  \dot{\psi
}\right]  .
\end{equation}

Recall that when $V_{D}\left(  \psi\right)  $ is zero, or constant, i.e.
$\alpha=-2$, the symmetry vector $X^{3}$ still exists.

\subsection{Brans-Dicke theory with $\omega_{BD}\left(  \phi\right)  =0$}

In the special where the Brans-Dicke parameter is zero, there exist additional
symmetries for the dynamical system. For arbitrary potential function
$V\left(  \psi\right)  $ there exist the additional Noether symmetries
\[
Y^{1}=\frac{1}{a}\partial_{\phi}~,~Y^{2}=\frac{t}{a}\partial_{\phi}%
\]
with conservation laws%
\begin{equation}
\bar{I}^{1}\left(  Y^{1}\right)  =a\dot{a}~,~\bar{I}^{2}\left(  Y^{2}\right)
=2ta\dot{a}+a^{2}.
\end{equation}

For the zero potential function $V_{A}\left(  \psi\right)  $, we derive the
additional symmetry vectors
\[
Y^{3}=\frac{1}{a}\left(  \psi\partial_{\phi}+3\partial_{\psi}\right)  .
\]
where the corresponding conservation laws is defined as%
\begin{equation}
\bar{I}^{3}\left(  Y^{3}\right)  =a\psi\dot{a}+a^{2}\dot{\psi}.
\end{equation}

\subsection{Coupling function $\omega_{BD}\left(  \phi\right)  =\frac{3}%
{2}\frac{\omega_{0}\phi}{\omega_{0}\phi+\phi_{0}}$}

For the coupling function $\omega_{BD}\left(  \phi\right)  =\frac{3}{2}%
\frac{\omega_{0}\phi}{\omega_{0}\phi+\phi_{0}}$, and for arbitrary potential
function $V\left(  \psi\right)  $, there exists the additional symmetry
vector
\begin{equation}
X^{4}=\frac{\sqrt{\omega_{0}\phi+\phi_{0}}}{a}\partial_{\phi},
\end{equation}
where from Noether's second theorem we calculate the conservation law%
\begin{equation}
I^{4}\left(  X^{4}\right)  =3a\sqrt{\omega_{0}\phi+\phi_{0}}\dot{a}.
\end{equation}
This is the only case calculated before in \cite{com1} for the constant
potential $V\left(  \psi\right)  =V_{0}$. But as we can see from the present
analysis conservation law $I^{4}\left(  X^{4}\right)  $ exists for any
potential function $V\left(  \psi\right)  $.

The results of the symmetry classification scheme are summarized in Table
\ref{tabl1}.%

\begin{table}[tbp] \centering
\caption{Noether symmetry classifcation for the generalized Brans-Dicke cosmological model.}%
\begin{tabular}
[c]{cccccc}\hline\hline
$\mathbf{\omega}\left(  \mathbf{\phi}\right)  $%
$\backslash$%
$\mathbf{V}\left(  \mathbf{\psi}\right)  $ & $\mathbf{V}\left(  \mathbf{\psi
}\right)  $ & $\mathbf{Zero}$ & $\mathbf{V}_{0}$ & $\mathbf{V}_{0}%
\mathbf{e}^{-\lambda\psi}$ & $\mathbf{V}_{0}\left(  \mathbf{\psi-\psi}%
_{0}\right)  ^{2+\alpha}$\\\hline
$\mathbf{V}\left(  \mathbf{\psi}\right)  $ & $X^{0}$ & $X^{0},X^{1},X^{2}$ &
$X^{0},X^{1}$ & $X^{0},X^{1+2}$ & $X^{0}$\\
$\mathbf{\omega}_{BD}^{0}\neq0$ & $X^{0}$ & $X^{0},X^{1},X^{2},X^{3}$ &
$X^{0},X^{1},X^{3}$ & $X^{0},X^{1+2}$ & $X^{0},X^{3}$\\
$\mathbf{\omega}_{BD}^{0}=0$ & $X^{0},Y^{1},Y^{2}$ & $X^{0},X^{1},X^{2}%
,X^{3},Y^{1},Y^{2},Y^{3}$ & $X^{0},X^{1},X^{3},Y^{1},Y^{2}$ & $X^{0},X^{1+2}$
& $X^{0},X^{3},Y^{1},Y^{2}$\\
$\frac{3}{2}\frac{\mathbf{\omega}_{0}\mathbf{\phi}}{\mathbf{\omega}%
_{0}\mathbf{\phi}+\phi_{0}}$ & $X^{0}$ & $X^{0},X^{1},X^{2},X^{4}$ &
$X^{0},X^{1},X^{4}$ & $X^{0},X^{1+2},X^{4}$ & $X^{0}$\\\hline\hline
\end{tabular}
\label{tabl1}%
\end{table}%

\section{New integrable models via the Noether symmetry analysis}

\label{sec4}

The gravitational field equations form a Hamiltonian system of dimension
three, which means that in order to determine the integrability, we need to
derive three conservation laws that are independent and in involution. The
constraint equation always exists; however, not all the functions derived by
the Noether classification scheme lead to an integrable dynamical system. For
the Brans-Dicke theory and the power-law potential, the integrability property
was found previously by deriving hidden symmetries that lead to the
construction of quadratic in the momentum conservation laws. Note that in this
study, we investigated point symmetries that lead to linear in the momentum
conservation laws.

In the following lines, we proceed with the derivation of a solution for the
exponential potential~$V_{C}\left(  \psi\right)  =V_{0}e^{-\lambda\psi}$, in
the case where $\omega_{BD}\left(  \phi\right)  =0$, and $\omega_{BD}=\frac
{3}{2}\frac{\omega_{0}\phi}{\omega_{0}\phi+\phi_{0}}$.

\subsection{Brans-Dicke theory with $\omega_{BD}\left(  \phi\right)  =0$}

In the case with $\omega_{BD}\left(  \phi\right)  =0$, the gravitational field
equations read%
\begin{equation}
3\phi H^{2}+3H\dot{\phi}+\frac{1}{2}\dot{\psi}^{2}+V\left(  \psi\right)  =0,
\end{equation}%
\begin{equation}
\dot{H}+2H^{2}=0,
\end{equation}%
\begin{equation}
2\left(  2\dot{H}+3H^{2}\right)  \phi+2\ddot{\phi}+4H\dot{\phi}-\dot{\psi}%
^{2}+2V\left(  \psi\right)  =0,
\end{equation}%
\begin{equation}
\ddot{\psi}+3H\dot{\psi}+V_{,\psi}\left(  \psi\right)  =0.
\end{equation}
Therefore the scale factor is always $a\left(  t\right)  =a_{0}\sqrt{t}$, i.e.
$H\left(  t\right)  =\frac{1}{2t}$, which corresponds to the radiation universe.

For the exponential potential there exist the conservation law $I^{1+2}\left(
X^{1+2}\right)  $ and the constraint equation, which lead to the reduced
system%
\begin{align}
\frac{I^{1+2}}{3\sqrt{t}}-\left(  \phi+t\left(  \dot{\phi}+2\frac{\dot{\psi}%
}{\lambda}\right)  \right)   &  =0,\\
\frac{3}{2t}\phi+2V_{0}e^{-\lambda\psi}+3\dot{\phi}+t\dot{\psi}^{2}  &  =0.
\end{align}

\subsection{Coupling function $\omega_{BD}\left(  \phi\right)  =\frac{3}%
{2}\frac{\omega_{0}\phi}{\omega_{0}\phi+\phi_{0}}$}

For the exponential potential $V_{C}\left(  \psi\right)  $ and the coupling
function $\omega_{BD}\left(  \phi\right)  =\frac{3}{2}\frac{\omega_{0}\phi
}{\omega_{0}\phi+\phi_{0}}$, we apply the change of variable $\phi
=-4\frac{\phi_{0}}{\omega_{0}}+\frac{^{\varphi2}}{4a^{2}}$ and the point-like
Lagrangian function (\ref{bd.07}) becomes%
\begin{equation}
L\left(  a,\dot{a},z,\dot{z},\psi,\dot{\psi}\right)  =3\frac{\phi_{0}}%
{\omega_{0}}a\dot{a}^{2}-\frac{1}{2}a^{3}\dot{\psi}^{2}-\frac{3}{4}%
a\dot{\varphi}^{2}+a^{3}V_{0}e^{-\lambda\psi}\text{.} \label{sd1}%
\end{equation}
Without loss of generality, we can assume $\phi_{0}=\omega_{0}$. In these
variables, the cosmological model is equivalent to that of a quintessence
scalar field with an exponential potential and a radiation fluid source. The
analytic solution for this cosmological model has been derived previously in
\cite{gend}, so we omit the presentation of the solution.

\subsubsection{Solution for arbitrary potential}

Nevertheless, for arbitrary potential $V\left(  \psi\right)  $, the point-like
Lagrangian (\ref{sd1}) reads%
\begin{equation}
L\left(  a,\dot{a},z,\dot{z},\psi,\dot{\psi}\right)  =3a\dot{a}^{2}-\frac
{1}{2}a^{3}\dot{\psi}^{2}-\frac{3}{4}a\dot{\varphi}^{2}+a^{3}V\left(
\psi\right)  \text{,}%
\end{equation}
which is that of a quintessence scalar field with an arbitrary potential and a
radiation fluid source. Indeed, this model is integrable via nonlocal
transformations, as found in \cite{gend}.

Indeed, conservation law, $I^{4}$ becomes $I^{4}\simeq a\dot{\varphi}$. By
replacing the conservation law we end with a reduced Hamiltonian system
described by the Lagrangian function%
\begin{equation}
L\left(  a,\dot{a},z,\dot{z},\psi,\dot{\psi}\right)  =3a\dot{a}^{2}-\frac
{1}{2}a^{3}\dot{\psi}^{2}+\frac{\hat{I}^{4}}{a}+a^{3}V\left(  \psi\right)  ,
\end{equation}
where $\hat{I}^{4}=-\frac{3}{4}I^{4}$.

The solution of the latter gravitational model is \cite{gend}
\begin{equation}
\phi(\tau)=\pm\frac{\sqrt{6}}{6}\int\left[  \left(  F^{\prime}\left(
\tau\right)  -6\gamma\rho_{m0}e^{F\left(  \tau\right)  -\frac{2}{3}\tau
}\right)  \right]  ^{1/2}\!d\tau,
\end{equation}
where%
\begin{equation}
V(\tau)=\frac{1}{12}e^{-F(\tau)}\left(  1-F^{\prime}(\tau)\right)  +\,\hat
{I}^{4}\,e^{-\frac{2}{3}\tau}\label{so.05}%
\end{equation}
where we have introduced the change of variable $dt=e^{\frac{1}{2}F\left(
\tau\right)  }d\tau$, and $a\left(  \tau\right)  =\frac{\tau}{3}$, with Hubble
function $H\left(  \tau\right)  =\frac{1}{3}e^{-\frac{1}{2}F\left(
\tau\right)  }$, thus the FLRW line element reads
\begin{equation}
ds^{2}=-e^{F\left(  \tau\right)  }d\tau^{2}+e^{\tau/3}(dx^{2}+dy^{2}%
+dz^{2}).\label{SF.12}%
\end{equation}

Hence, for any function $F\left(  \tau\right)  $ we can reconstruct a
corresponding potential function $V\left(  \psi\right)  $, such that the
theory to be cosmological viable. 

\section{Conclusions}

\label{sec5}

In this study, we solved the complete symmetry classification problem for a
Hamiltonian model that describes a constraint cosmological model with two
scalar fields, one minimally coupled to gravity and another nonminimally
coupled to gravity. The gravitational model under our consideration belongs to
the family of generalized Brans-Dicke theory. The evolution of the physical
parameters depends on two arbitrary functions: the coupling function of the
generalized Brans-Dicke field with the gravitational field, and the scalar
field potential which gives the mass of the minimally coupled scalar field.

We applied Noether's first theorem and required the existence of variational
symmetries for this dynamical system. The symmetry condition leads to a system
of linear inhomogeneous partial differential equations where the free
functions of the theory are constrained. The results of the classification
problem are presented in Table \ref{tabl1}.

By using Noether's second theorem, we constructed the corresponding
conservation laws and discussed the integrability properties of the field
equations. Finally, we were able to reduce the given cosmological model to the
equivalent form of another theory and apply previous results to solve the
gravitational field equations.

This work completes the results presented in \cite{com1}. More specifically,
from the sixteen different cases of the Noether symmetry classification scheme
identified in this work, in \cite{com1} only a fraction of the whole
classification was presented.

The symmetry classification problem of differential equations is a classical
problem of applied mathematics. Nevertheless, nowadays, Noether symmetry
analysis has been used in gravitational physics, but often not in a rigorous
mathematical formalism, which has resulted in new researchers producing
various inaccurate results or incomplete studies.

\begin{acknowledgments}
The author thanks the support of Vicerrector\'{\i}a de Investigaci\'{o}n y
Desarrollo Tecnol\'{o}gico (Vridt) at Universidad Cat\'{o}lica del Norte
through N\'{u}cleo de Investigaci\'{o}n Geometr\'{\i}a Diferencial y
Aplicaciones, Resoluci\'{o}n Vridt No - 096/2022 and Resoluci\'{o}n Vridt No - 098/2022.
\end{acknowledgments}

\appendix

\section{Symmetry conditions}

\label{app1}

For the Lagrangian function (\ref{bd.07}) the Noether symmetry condition
(\ref{Lie.5}) provides the following system of partial differential equations%

\[
0=\left(  \phi\eta_{a}+a\eta_{\phi}+2a\phi\left(  \frac{\partial\eta_{a}%
}{\partial a}+a\frac{\partial\eta_{\phi}}{\partial a}\right)  \right)
-\frac{\partial\xi}{\partial t}a\phi,
\]%
\[
0=3\left(  2\phi^{2}\eta_{a}+a\eta_{\phi}\right)  +3a\phi^{2}\left(
\frac{\partial\eta_{a}}{\partial a}+\frac{\omega_{BD}}{3\phi^{2}}a^{2}%
\frac{\partial\eta_{\phi}}{\partial a}\right)  +6\phi^{2}\left(
\frac{\partial\eta_{a}}{\partial\phi}+3a\frac{\partial\eta_{\phi}}%
{\partial\phi}\right)  -3a\phi\frac{\partial\xi}{\partial t},
\]%
\[
0=\frac{\omega_{BD}}{\phi}\left(  3\phi\eta_{a}-a\eta_{\phi}\right)
+a\eta_{\phi}\frac{\partial\omega_{BD}}{\partial\phi}+6\phi^{2}\frac
{\partial\eta_{a}}{\partial\phi}+2a\frac{\omega_{BD}}{\phi}\frac{\partial
\eta_{\phi}}{\partial\phi}-\omega_{BD}a\frac{\partial\xi}{\partial t},
\]%
\[
0=a^{2}\frac{\partial\eta_{\psi}}{\partial a}+3a\phi\frac{\partial}%
{\partial\psi}\left(  2\eta_{a}+a\eta_{\phi}\right)  ,
\]%
\[
0=a\frac{\partial\eta_{\psi}}{\partial\phi}+\frac{\partial}{\partial\psi
}\left(  3\phi^{2}\eta_{1}+a\omega_{BD}\eta_{\phi}\right)  ,
\]%
\[
0=3\eta_{a}+2a\frac{\partial\eta_{\psi}}{\partial\psi}-a\frac{\partial\xi
}{\partial t},
\]%
\[
0=a^{2}\left(  3\eta_{a}+a\eta_{\psi}\frac{\partial V}{\partial\psi}%
+aV\frac{\partial\xi}{\partial t}\right)  +f_{,t},
\]%
\[
0=3a\left(  2\eta_{a}+a\phi\eta_{\phi}\right)  +f_{,a},
\]%
\[
0=\frac{a^{2}}{\phi}\left(  3\phi^{2}\eta_{a}+a\eta_{\phi}\right)  +f_{,\phi},
\]%
\[
0=a^{3}\eta_{\psi}+f_{,\psi},
\]%
\[
\xi=\xi\left(  t\right)  .
\]

The solution of the latter system constraint the functional forms for the
coefficients $\xi,~\eta_{a},~\eta_{\phi}$ and $\eta_{\psi}$ of the vector
field $X$, and the free functions $\omega_{BD}\left(  \phi\right)  $ and
$V\left(  \phi\right)  $ of the Brans-Dicke gravitational model.

\end{document}